\documentclass{aa}  

\usepackage{graphicx}
\usepackage{txfonts}
\usepackage{subcaption}
\usepackage{hyperref}
\begin{document}

   \title{Cosmological Inference with Cosmic Voids and Neural Network Emulators}

   \author{Kai Lehman
          \inst{1,2}\thanks{Corresponding author; \email{kai.lehman@physik.lmu.de}}
          \and
          Nico Schuster
          \inst{1,3}
          \and
          Luisa Lucie-Smith
          \inst{4,5}
          \and
          Nico Hamaus
          \inst{1,2}
          \and
          Christopher T. Davies
          \inst{1}
          \and
          Klaus Dolag
          \inst{1,4}
          }

   \institute{Universit\"ats-Sternwarte M\"unchen, Fakult\"at f\"ur Physik, Ludwig-Maximilians-Universit\"at, Scheinerstr. 1, 81679 M\"unchen, Germany 
         \and
             Excellence Cluster ORIGINS, Boltzmannstr. 2, 85748 Garching, Germany 
         \and
              Aix-Marseille Universit\'e, CNRS/IN2P3, CPPM, Marseille, France
         \and
             Max-Planck-Institut für Astrophysik, Karl-Schwarzschild-Str. 1, 85748 Garching, Germany
         \and
             Universit\"at Hamburg, Hamburger Sternwarte, Gojenbergsweg 112, 21029 Hamburg, Germany
             }

   \date{Received; accepted}

  \abstract
   {Cosmic Voids are a promising probe of cosmology for spectroscopic galaxy surveys due to their unique response to cosmological parameters. Their combination with other probes promises to break parameter degeneracies.}
   {Due to simplifying assumptions, analytical models for void statistics are only representative of a subset of the full void population. We present a set of neural-based emulators for void summary statistics of watershed voids, which retain more information about the full void population than simplified analytical models.}
   {We build emulators for the void size function and void density profiles traced by the halo number density using the \textsc{Quijote} suite of simulations for a broad range of the $\Lambda\mathrm{CDM}$ parameter space. The emulators replace the computation of these statistics from computationally expensive cosmological simulations. We demonstrate the cosmological constraining power of voids using our emulators, which offer orders-of-magnitude acceleration in parameter estimation, capture more cosmological information compared to analytic models, and produce more realistic posteriors compared to Fisher forecasts.}
   {We find that the parameters $\Omega_m$ and $\sigma_8$ in this \textsc{Quijote} setup can be recovered to 14.4\% and 8.4\% accuracy respectively using void density profiles; including the additional information in the void size function improves the accuracy on $\sigma_8$ to 6.8\%. We demonstrate the robustness of our approach to two important variables in the underlying simulations, the resolution, and the inclusion of baryons. We find that our pipeline is robust to variations in resolution, and we show that the posteriors derived from the emulated void statistics are unaffected by the inclusion of baryons with the \texttt{Magneticum} hydrodynamic simulations. This opens up the possibility of a baryon-independent probe of the large-scale structure.}
   {}


   \maketitle
%

\section{Introduction}

Cosmological surveys of the past decade~\citep{hetdex_08, boss_13, kids_13,sptsz_15, hsc_15, des_16, eboss_16} and the future~\citep{lsst_09,euclid_11,4most_12,desi_13,pfs_14,ska_2015,roman_15,erosita_21} map the large-scale structure of the universe with unprecedented statistical power, which has enabled the precise inference of cosmological parameters.
These modern surveys are able to measure the distribution of galaxies with very high accuracy, which has allowed for the use of cosmic voids as cosmological probes~\citep[e.g.,][]{contarini_2022,hamaus_2022,bonici_2023,radinovic_2023,song_2025}. For a robust cosmological analysis of voids, survey depth and completeness of galaxy catalogs is important as to not falsely identify regions of unobserved galaxies as voids. At the same time, their size is on the scale of tens of Megaparsec, and they make up the majority of the volume of the universe~\citep{sheth_2004}, meaning a large survey volume is also required. Due to these limitations, their study has only recently become feasible with the advent of precision cosmology~\citep[see][for a recent review]{pisani_2019,moresco_2022}.

The Baryon Oscillation Spectroscopic Survey (BOSS)~\citep{dawson_2013}, measuring spectroscopic redshifts of 1.5 million galaxies, has shown that such surveys are ideal for void studies. Statistics describing the observed BOSS void populations have been shown to be useful for cosmological parameter inference, including, for example their abundance \citep{contarini_2023,song_2025}, the Alcock-Paczyński test \citep{alcock_1979,hamaus_2016,hamaus_2020} or the void probability function~\citep{garcia_2024}. An especially promising property of cosmic voids currently sparking further investigation is their complementary response to cosmological parameters with other probes of the large-scale structure. For example, the well-known $\Omega_m$-$\sigma_8$ degeneracy, typically obtained from 3x2pt analyses or cluster number counts, exhibits an orthogonal correlation direction when extracted from the void size function (VSF; \citealt{pelliciari_2023,contarini_2023}). More generally, voids are complementary to overdense structures like halos, and a combination of the two probes is expected to break crucial parameter degeneracies~\citep{kreisch_2022}. 

These analyses are very promising and while it has been shown that some of the fundamental void properties follow simpler models~\citep{hamaus_2014, stopyra_2021, schuster_2023,schuster_2024}, the fidelity of void models is limited in analytical calculations. There are currently no theoretical or analytical models that accurately describe the void density profiles including the response to changes in cosmological parameters, and the VSF of watershed voids does not accurately follow theoretical models unless a simplifying ``cleaning'' procedure is applied~\citep{sheth_2004, jennings_2013, ronconi_2017, contarini_2023}. On the other hand, more sophisticated models are readily available in the form of $N$-body simulations, which simulate the entire non-linear density field including the full void population. The limiting factor in this case is the prohibitive cost of such simulations for standard MCMC analyses. This is where machine learning methods are gaining traction in cosmology. The feasibility of inference is no longer bound to fast analytical models. More expensive simulations and therefore more intricate models can be coupled with machine learning, as it accelerates inference by many orders of magnitude. This is possible, for example, with the use of emulators \citep[e.g.][for examples in cosmology]{knabenhans_2019,mcclintock_2019,zhai_2019,knabenhans_2021,mancini_2022,gong_2023,bai_2024,storey_2024}, more recently also on the field-level~\citep{ramanah_2020,kaushal_2022,doeser_2023,jamieson_2023,jamieson_2024,ibanez_2024}. Other examples exploiting the computational efficiency of machine learning in cosmological and astrophysical settings are photometric redshift estimation with neural networks \citep[e.g.,][]{hoyle_2015,schuldt_2021} or automatic detection of gravitational lenses \citep[e.g.,][]{wilde_2022}. For voids, classifiers have been used in~\citep{cousinou_2019} to help remove Poisson noise in void catalogs.

There has been some exploratory work making use of machine learning methods to investigate cosmic voids for cosmological inference. \citet{kreisch_2022} present a large catalog of voids in the \textsc{Quijote} simulations with the \textsc{vide} void finder \citep{sutter_2015}. Using Fisher forecasts, they demonstrate how voids provide complementary cosmological information to that from halos and explore the constraining power of the void size function using a moment network~\citep{jeffrey_2020}. This catalog is also used by \citet{wang_2022}, who show how machine learning can be used to extract cosmological information from the properties of cosmic voids. In their work they use fully connected neural networks to infer cosmological parameters from ellipticity and density contrast distributions of voids in the simulations. Furthermore, they perform inference with void catalogs directly using permutation invariant neural network architectures known as deep set networks \citep{zaheer2017deep}. With this likelihood-free approach they are able to constrain $\Omega_m$, $\sigma_8$ and $n_s$. In another work, \citet{thiele_2023} provide a constraint on the neutrino mass from voids with implicit likelihood inference where they propose that voids mostly put a lower bound on the sum of neutrino masses. Moreover, the analysis of~\citep{Wang_2024} uses likelihood-free inference to learn about cosmology using individual galaxies inside voids and~\citep{fraser_2024} model the void-galaxy cross correlation function with an emulator-based approach.

In this work, we build upon the Fisher forecasts presented in \citet{kreisch_2022} and derive non-Gaussian posteriors for the cosmological parameters using neural network emulators of a range of void properties. We further explore the robustness of these statistics with respect to simulation specifications, such as resolution and the inclusion of baryons. All emulators are trained on the latin-hypercube $\Lambda$CDM \textsc{Quijote} set of simulations. 

This paper is structured as follows: In Section~\ref{sec:simulations} we present the simulations used for the training and testing of the statistical models, and the void definition used in this work as given by the VIDE void finder. In Section~\ref{sec:emulators} we train and test the neural network emulators, which subsequently allows us to quantify the response of void statistics to cosmological parameters and hence to forecast cosmological constraints with cosmic voids on both the \textsc{Quijote} and \texttt{Magneticum} simulations (Section~\ref{sec:results}). Finally, our conclusions are presented in Section~\ref{sec:conclusions}.

\section{Simulations \& Void Finder} \label{sec:simulations}

\subsection{The \textsc{Quijote} Simulations} \label{sec:quijote}

The \textsc{Quijote} suite of gravity-only $N$-body simulations~\citep{navarro_2020} contains over 44100 boxes with more than 7000 combinations of cosmological parameters. The fiducial resolution boxes have $512^3$ particles in a volume of $1(h^{-1}\mathrm{Gpc})^3$. The cosmological parameters that are varied in the $\Lambda$CDM simulations are $\Omega_m$ and $\Omega_b$, $h$, $n_s$ and $\sigma_8$. We make use of the friends-of-friends halo catalogs at redshift $z=0$ to identify the voids. The simulations considered here consist of the following two sets:

\begin{itemize}
    \item Cosmic Variance (CV): In this set, all cosmological parameters are fixed to their fiducial values from~\citet{planck_2018} and only the initial conditions are varied. This allows for the reliable estimation of covariance matrices, which quantify the impact of cosmic variance and correlations in void statistics. We use a subset of 1000 simulations from this set to estimate covariance matrices.
    \item Latin Hypercube (LH): we use the simulations from this set to train the machine learning models, as both the cosmological parameters \emph{and} the initial conditions are varied. The cosmological parameters are varied according to a latin-hypercube sampling scheme. In total this set consists of 2000 simulations.
\end{itemize}

The cosmological parameters of all of the simulations described above are given in Table \ref{tab:sim_params}. We use halo catalogs generated with the friends-of-friends~\citep{davis_1985} algorithm with the linking length parameter $b=0.2$ and a minimum dark matter particle number of $20$.
\begin{table*}
    \caption{Cosmological parameters in the \textsc{Quijote} $\Lambda$CDM and the \texttt{Magneticum} simulations. All simulations assume flat curvature.}
\label{tab:sim_params}      
\centering          
    \begin{tabular}{c c c c c c}
\hline\hline       
          & $\Omega_m$ & $\Omega_b$ & $h$ & $n_s$ & $\sigma_8$ \\
        \hline
        \textsc{Quijote} CV & 0.3175 & 0.049 & 0.6711 & 0.9624 & 0.834 \\
        \textsc{Quijote} LH upper bound & 0.5 & 0.07 & 0.9 & 1.2 & 1.0\\
        \textsc{Quijote} LH lower bound & 0.1 & 0.03 & 0.5 & 0.8 & 0.6\\
        \texttt{Magneticum} & 0.272 & 0.0456 & 0.704 & 0.963 & 0.809 \\
\hline                    
\end{tabular}
\end{table*}

\subsection{The Magneticum Simulations}
Hydrodynamic simulations also model baryonic effects, including feedback effects originating from stars and active galactic nuclei (AGN). For example supernovae or jets from AGNs can significantly redistribute gas and alter the matter density field on scales up to a few Megaparsec~\citep{rudd_2008}. It therefore appears unlikely at first glance that a statistical model trained on a gravity-only simulation might work on a hydrodynamical simulation. However, cosmic voids are pristine environments where baryonic effects have very limited impact on both their density profiles and their abundance~\citep{paillas_2017,schuster_2024}. If this is true, it may be that our neural network emulators are able to generalize to hydrodynamical simulations, even if trained solely on dark-matter-only simulations. Moreover, numerical modeling of baryonic feedback is highly uncertain, as their effect on the matter distribution differs in a variety of simulations~\citep{chisari_2019}. In light of this modeling uncertainty, a robust probe of the large-scale structure would be desirable. In order to test whether voids are suitable as such, we perform mock inference with the state-of-the-art hydrodynamic \texttt{Magneticum}\footnote{http://www.magneticum.org/} simulation (box 0). The \texttt{Magneticum} simulation models complex baryonic physical processes such as feedback mechanisms and radiative cooling; these simulations have been used to study the Sunyeav-Zel'dovich power spectrum \citep{dolag_2016b}, multiple aspects of galaxy clusters \citep[e.g.][]{gupta_2017,biffi_2013,biffi_2022} or properties of galaxies \citep[e.g.][]{hirschmann_2014,lotz_2021}. They are based on the tree particle mesh smoothed particle hydrodynamics code \textsc{P-gadget3}, an improved version of \textsc{P-gadget2}~\citep{springel_2005} with the SPH solver from~\citet{beck_2016}. Naturally, these mechanisms affect structure formation in the simulation. The simulations adopt a flat $\Lambda$CDM cosmology with parameters from WMAP~\citep{komatsu_2011}. A summary of the cosmological parameters used in the \textsc{Quijote} and \texttt{Magneticum} simulations is shown in Table~\ref{tab:sim_params}.

In this analysis we use box 0 from the \texttt{Magneticum} simulations, which has been used in previous studies on cosmic voids~\citep{pollina_2017,schuster_2023,schuster_2024}. The simulation evolves $2\times4536^3$ particles within a cube with a side length of $2688\, h^{-1}\mathrm{Mpc}$. Compared to the \textsc{Quijote} simulations, it is clear that the \texttt{Magneticum} simulation has a much higher resolution. This is important to take into account, as void statistics are sensitive to the density of tracers in which they are identified. A higher tracer density, for example, allows us to resolve smaller voids, but fragments larger voids when neglecting merged super-structures \citep{schuster_2023}. This then needs to be disentangled from cosmological effects. Additionally, when comparing box sizes, box 0 is much larger than the \textsc{Quijote} simulations. This in turn yields lower cosmic variance in \texttt{Magneticum} compared to \textsc{Quijote} when averaging over the entire available volume in both cases. In order to address these two issues, we cut small halos from the \texttt{Magneticum} halo catalog to achieve comparable halo densities and consider only a $1h^{-1}\mathrm{Gpc}$ sub-box to match the simulation volumes. We provide more detail on how we apply the emulator on \texttt{Magneticum} boxes in Section~\ref{sec:other_sims}.

\subsection{VIDE}

In order to identify voids in the simulations, we need to choose a void finder and a corresponding void definition. In this analysis we use VIDE\footnote{\url{https://bitbucket.org/cosmicvoids/vide_public/}}  \citep{sutter_2015}, an extension of the ZOBOV~\citep{neyrinck_2008} algorithm. VIDE identifies voids in a topological manner based on the watershed transform~\citep{platen_2007}. Voids are hence defined as the regions enclosed by local maxima in the density field, which define the void boundary. As voids can't be directly observed in the dark matter field, we need to estimate them with a tracer population (e.g. galaxies or halos). Within VIDE this is achieved by defining Voronoi cells for each tracer, which assigns densities inversely proportional to their volume. By definition of the watershed algorithm, each void is associated with a local minimum, which is a cell of lower density than any of its neighbors. Then the regions belonging to a given void are identified by assigning each tracer particle to its lowest density neighbor until a minimum is reached. Intuitively, this procedure is analogous to the filling of water basins on a non-uniform surface. A droplet of water anywhere on this surface will flow towards the minimum associated with a given void. For each void the effective void radius $r_\mathrm{v}$ is calculated, which is defined as the radius of a sphere with the corresponding volume of the void $V_\mathrm{v}$:

\begin{equation}
    r_\mathrm{v} = \left(\frac{3}{4\pi}V_\mathrm{v}\right)^{1/3},
\end{equation}

where the void volume is the sum of all member Voronoi cells. We note that even though we use the void radius to characterize voids, they are not actually spherical and can take on very complicated shapes. We identify all voids in the halo field.

\section{Cosmology with Emulators} \label{sec:emulators}

In this section, we introduce the neural network emulators for void statistics and then describe the cosmological parameter inference pipeline.

\begin{figure*}
     \centering
     \begin{subfigure}[b]{.9\textwidth}
         \centering
        \includegraphics[width=\textwidth]{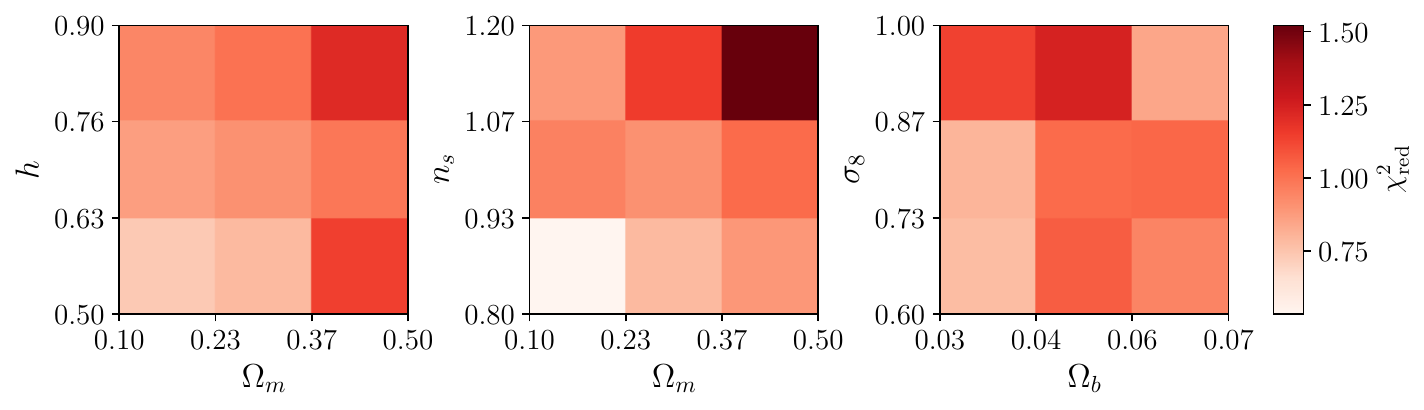}
        \caption{VSF emulator}
     \end{subfigure}
     \hfill
     \begin{subfigure}[b]{.9\textwidth}
         \centering
        \includegraphics[width=\textwidth]{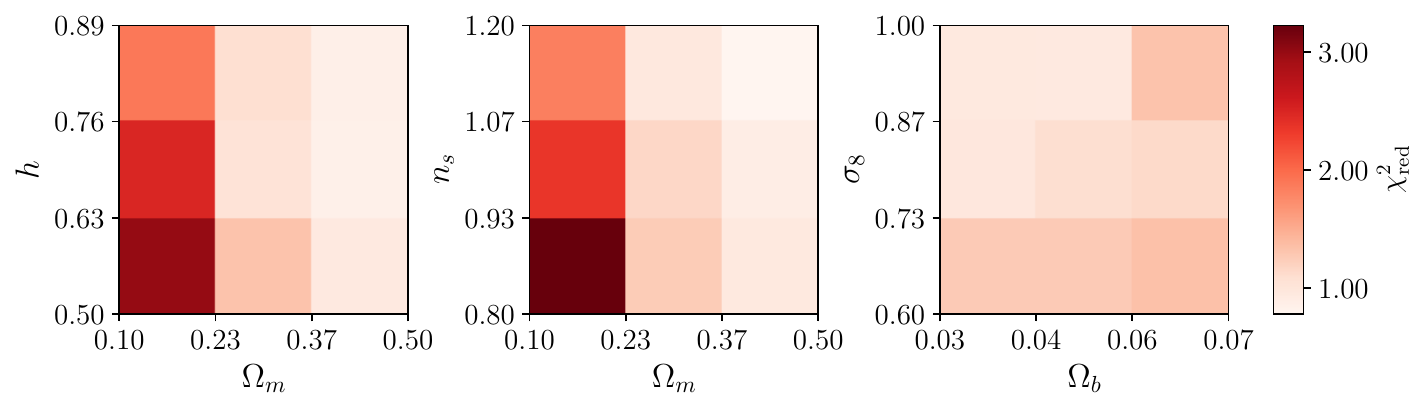}
        \caption{Void Density Profiles (VP) emulator}
        \label{fig:test_den}
     \end{subfigure}
    \caption{Emulator accuracy across the cosmological parameter space, represented by the median reduced chi-squared ($\chi^2_\mathrm{red}$) values of the test set predictions across 2D cosmological parameter spaces. The $\chi^2_\mathrm{red}$ of the void size function are shown in the upper panels; those of the void density profiles in the lower panels. The void size function emulator is accurate over the entire parameter space, while the void density profile emulator is accurate with the exception of low $\Omega_m$ values.}
    \label{fig:chis}
\end{figure*}

\subsection{Neural Network Emulators for Void Statistics}
We train two distinct emulators: one for the void density profiles and one for the void size function. The emulators take as input the five cosmological parameters and return the void density profiles (or the void size function). The following steps are performed for the training process of each emulator:

\begin{enumerate}
    \item Training data collection: we start with measuring the void density profile and void size function from each simulation in the training data.
    
    \item Data pre-processing: neural networks tend to perform better for pre-processed, standardized data. This is especially the case for data vectors that span a wide range of numerical values. For subsequent parameter inference, these transformations have to be reversed. The cosmological parameters which serve as inputs $\theta$ are always standardized to zero mean and unit variance:
    \begin{equation}
        \theta_\mathrm{stand} = \frac{\theta-\mu_\theta}{\sigma_\theta},
    \end{equation}
    where $\theta_\mathrm{stand}$ is the standardized parameter and $\mu_\theta$ and $\sigma_\theta$ are the mean and the standard deviation of the parameter over the entire dataset. The dataset is split into 80\% training, 10\% validation and 10\% testing data.
    
    \item Emulator training: a neural network emulator is trained to predict a void statistic given (standardized) cosmological parameters. This is implemented using tensorflow \citep{abadi_2015}. All neural networks were trained on an NVIDIA GeForce GTX 1080 GPU. Neural networks contain two sets of parameters: hyperparameters, which must be manually set prior to the training process, and weights and bias parameters, which are instead learned by the network during training via stochastic gradient descent. Hyperparameters are scanned with a grid search such that the loss function is minimized on the validation set. This set is withheld in the optimization of the network's internal parameters.
    
    \item Emulator evaluation: the emulator is evaluated using the test set that has been withheld during both training and hyperparameter optimization. We conduct a twofold evaluation: on the one hand, we test whether the accuracy of the emulator is comparable in all regions of parameter space, i.e. if there are cosmological parameter values for which the emulator has poor accuracy. On the other hand, we also check whether there are particular parts of the data vector that have a larger prediction error. 
    We provide more details in Sec.~\ref{sec:trainingvalidation}.
\end{enumerate}

\subsubsection{Void Statistics Output}
\label{sec:trainingvalidation}

We calculate spherical number density profiles as a function of radius for each void around its volume-weighted barycenter by estimating the number density of halos in radial shells of width $0.1\times r_\mathrm{v}$. The radius of each shell is given in units of void radii of the corresponding void. The voids are grouped by radius into four size quartiles. All void density profiles in a given quartile are then stacked to produce a mean profile for that quartile. As the density estimation in the very underdense center of the void tends to be noisy~\citep{schuster_2023}, the innermost bin considered for our data-vector is the shell at $r=0.3\times  r_\mathrm{v}$. In order to further reduce the dimensionality, the outer bins are also cut from the data vector to obtain an array of length 25, putting the outermost bin considered at $r=2.7\times r_\mathrm{v}$. We expect the discarded bins to have little cosmological information, as the density profiles are expected to converge to the background density at far distances from void centers. We therefore prefer this approach to using larger bins  initially, which would lead to more information loss. This is done for each of the 2000 simulations in the \textsc{Quijote} LH set. The second statistic considered is the void size function (VSF), which we compute in $10$ bins from $10 h^{-1}\mathrm{Mpc}$ to $70 h^{-1}\mathrm{Mpc}$.

\subsubsection{Training Procedure}
We split the simulation budget into 80\% training, 10\% validation and 10\% testing data. While the training data is used to fit the parameters of the neural networks, the validation set is used for hyperparameter optimization and early stopping. The test set is only used once to evaluate the final models. For the density profile emulator we train a fully connected neural network to predict the stacked density profiles of all void radius quartiles at once. The output dimension is therefore $4\times 25 =100$. The VSF emulator is a separate neural network, which outputs the standardized logarithm of the void size function in 10 bins. 

As we are able to obtain a covariance matrix estimate from the CV \textsc{Quijote} simulations set, we use $\chi^2$ as the loss function for each model:
\begin{equation}
    \chi^2 = (\pmb t_\mathrm{pred} - \pmb t_\mathrm{sim}) \pmb C^{-1} (\pmb t_\mathrm{pred} - \pmb t_\mathrm{sim}),
\end{equation}
where $\pmb t_\mathrm{pred}$ is the network's prediction and $\pmb t_\mathrm{sim}$ is the data vector as measured in the simulation. $\pmb C^{-1}$ is the estimated precision matrix. We obtain this estimate by computing the density profiles and VSF in each of the 1000 CV simulations, see Section~\ref{sec:quijote}.

The best model for density profile emulation after hyperparameter optimization consists of 4 hidden layers with 512 neurons, all of which use the hyperbolic tangent activation function. We train with a batchsize of 1 and a learning rate of $10^{-4}$ using the Adam~\citep{kingma_2014} optimizer. The learning rate is further reduced by a factor of 0.1 if the validation loss does not improve after 60 epochs. We use early stopping with a very long patience of 120, but restore the weights of the best epoch after training. The best VSF model consists of 3 hidden layers with 256 neurons each. The network was also optimized using the Adam optimizer with an initial learning rate of $10^{-4}$ and a batch size of 1. The learning rate is decreased by a factor of 10 if the validation loss does not improve for 20 epochs. Early stopping is used if the validation loss does not improve for 40 epochs and the weights of the best performing model on the validation set are restored.

\subsubsection{Emulator Performance}
In order to examine the performance of the emulators, the loss function is computed for each sample in the test set separately. We then calculate $\chi^2_\mathrm{red}$ defined as 

\begin{equation}
    \chi^2_\mathrm{red} = \frac{\chi^2}{N_\mathrm{dof}},
\end{equation}

where $N_\mathrm{dof}$ is the number of degrees of freedom, i.e. in this case the number of output neurons of the neural network. 
This test is inspired by~\citet{gong_2023}, with the difference that they compare to a model prediction at the fiducial point. Due to the lack of an analytical model for voids, the comparison is done with the corresponding $N$-body simulation at each test node instead.
Having computed the loss for each test example individually, we project the parameter space along all parameters but 2. We then partition the remaining 2D parameter space into a 3x3 grid along the two remaining cosmological parameters and take the median of all test examples in each of the pixels. This is shown in Fig.~\ref{fig:chis}. The test allows us to identify whether the emulator's accuracy is similar over the whole parameter space, or if it depends on the value of the cosmological parameters.

The VSF emulator is of comparable performance over the entire parameter range. We find that at low values of $\Omega_m$ the performance of the density profile emulator drops off, likely due to larger variance in void statistics: smaller values of $\Omega_m$ at fixed $\sigma_8$ reduce the number of the small more numerous voids. This, in turn, results in noisier data vectors (and hence lower emulator accuracy). As the covariance is only estimated at the fiducial cosmology and its parameter dependence is neglected, this is not taken into account in the $\chi^2_\mathrm{red}$ calculation. The mean value over the test set for the VSF emulator is $\chi^2_\mathrm{red}=1.08$ and for the density profile emulator $\chi^2_\mathrm{red}=2.66$. The median of the density profile emulator is significantly lower at $\chi^2_\mathrm{red, median}=1.55$, which we attribute to the few outliers at low values of $\Omega_m$ that have also been detected in the previous test.

\begin{figure*}
     \centering
     \begin{subfigure}[b]{0.49\textwidth}
    \centering
    \includegraphics[width=\linewidth]{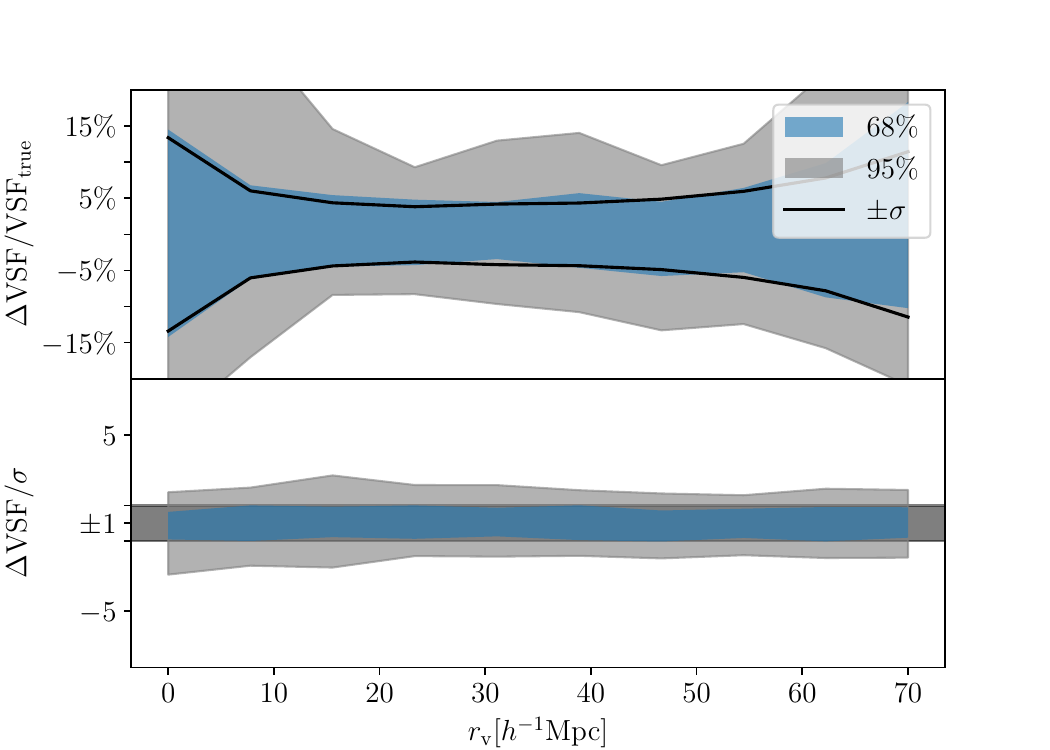}
    \caption{VSF}
    \label{fig:residuals_vsf}
     \end{subfigure}
     \hfill
     \begin{subfigure}[b]{0.49\textwidth}
    \centering
    \includegraphics[width=\linewidth]{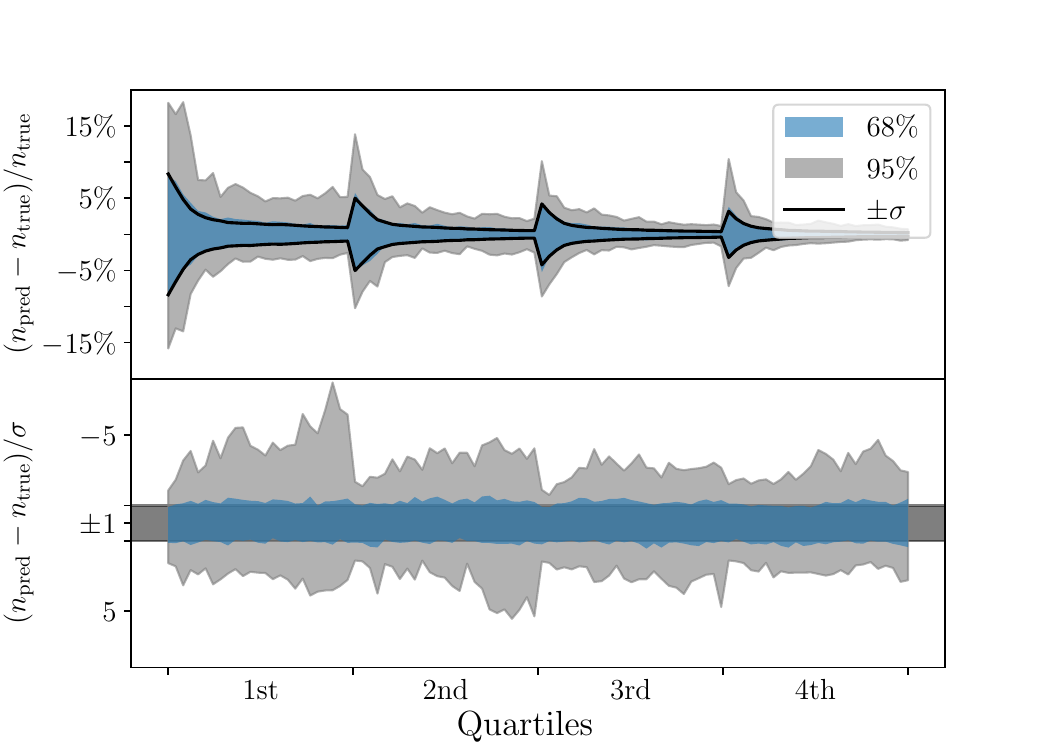}
    \caption{Density Profiles}
    \label{fig:residuals_den}
     \end{subfigure}
    \caption{Residuals of (a) the VSF emulator predictions and (b) the density profile emulator predictions on the test set show that both are highly accurate. The blue and gray bands indicate 68\% and 95\% of predictions on the test set, and the black line shows the deviation of $1\sigma$ as derived from the covariance estimate. For clarity we denote $\frac{\mathrm{d}n_\mathrm{v}}{\mathrm{d}\ln r_\mathrm{v}}[h^3\mathrm{Mpc}^{-3}]$ as $\mathrm{VSF}$ and $\Delta$ denotes the difference of its prediction with respect to the measurement from the corresponding box.}
    \label{fig:residuals}
\end{figure*}

We proceed with another way to quantify the performance of the emulator, this time directly in data space. We compute the residuals of the predictions of the test set and show them normalized by the true data vector, as well as by the standard deviation expected from cosmic variance in Figure \ref{fig:residuals}. We observe both emulators to be highly accurate with roughly 68\% of predictions within $1\sigma$. The 95\% contour for the density profile emulator is slightly larger than expected, which we trace back to the worse performance for low $\Omega_m$.

\subsection{Emulator Predictions}

\begin{figure*}
     \centering
     \begin{subfigure}[b]{0.49\textwidth}
    \centering
    \includegraphics[width=0.84\linewidth ]{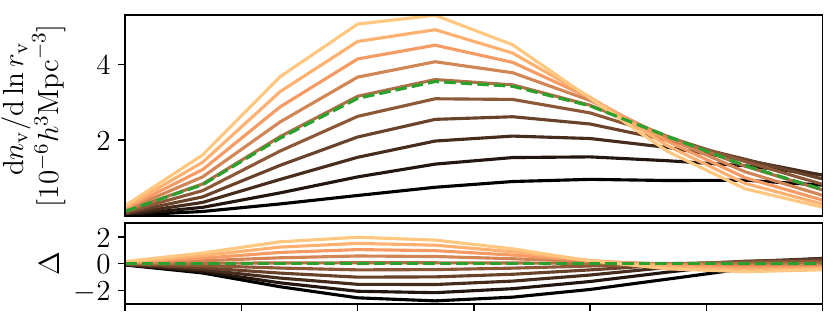}
     \end{subfigure}
     \begin{subfigure}[b]{0.49\textwidth}
    \centering
    \includegraphics[width=\linewidth]{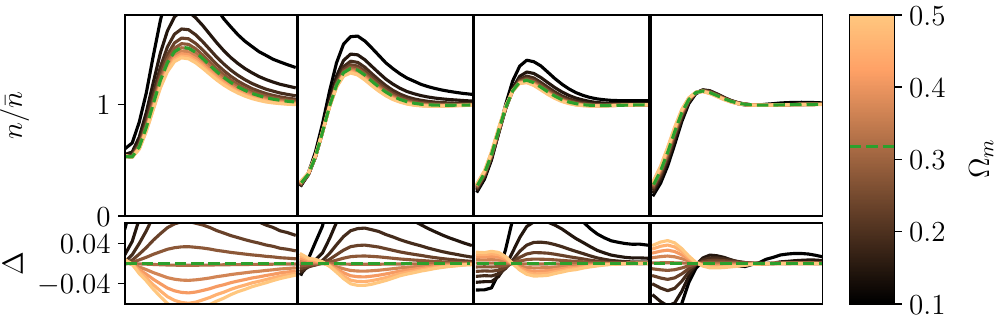}
     \end{subfigure}
     \hfill
     \begin{subfigure}[b]{0.49\textwidth}
    \centering
    \includegraphics[width=0.84\linewidth]{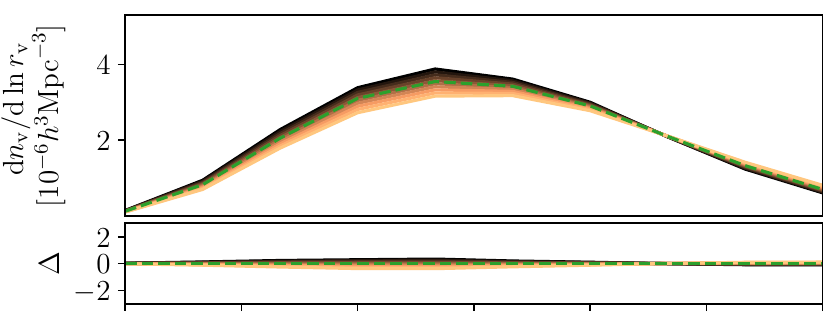}
     \end{subfigure}
     \begin{subfigure}[b]{0.49\textwidth}
    \centering
    \includegraphics[width=\linewidth]{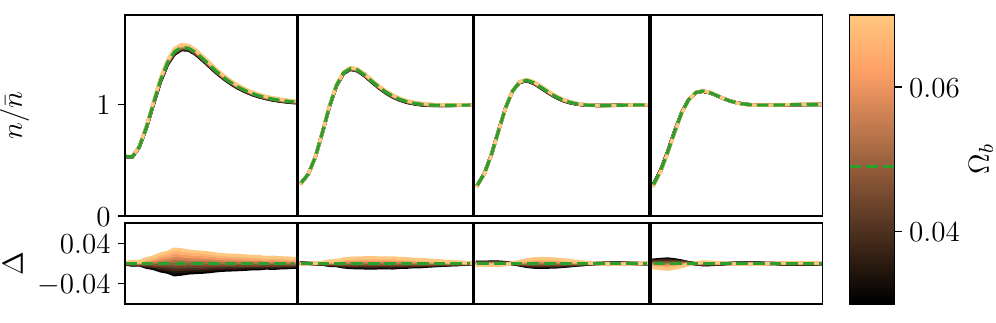}
     \end{subfigure}
     \begin{subfigure}[b]{0.49\textwidth}
    \centering
    \includegraphics[width=0.84\linewidth]{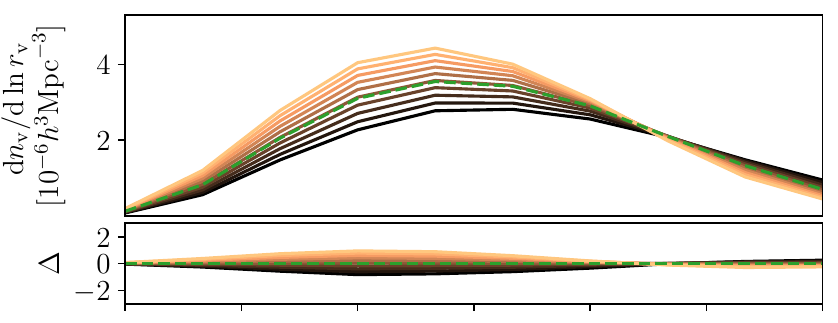}
     \end{subfigure}
     \begin{subfigure}[b]{0.49\textwidth}
    \centering
    \includegraphics[width=\linewidth]{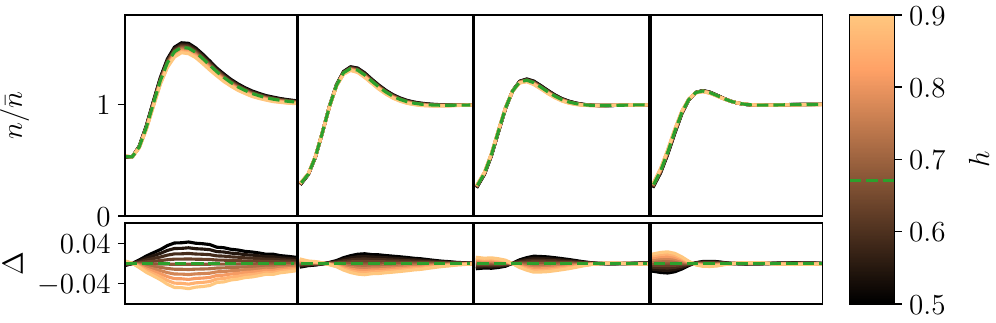}
     \end{subfigure}
     \begin{subfigure}[b]{0.49\textwidth}
    \centering
    \includegraphics[width=0.84\linewidth]{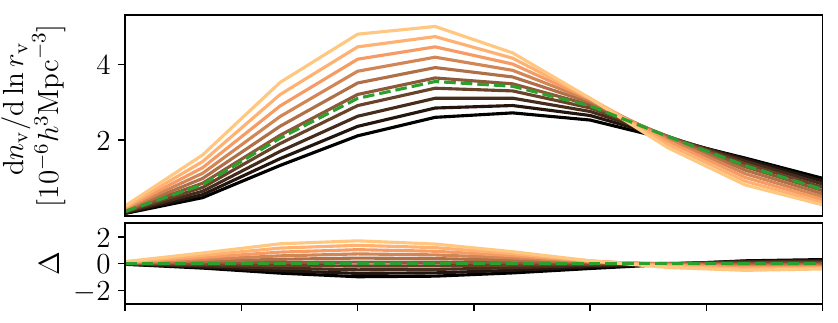}
     \end{subfigure}
     \begin{subfigure}[b]{0.49\textwidth}
    \centering
    \includegraphics[width=\linewidth]{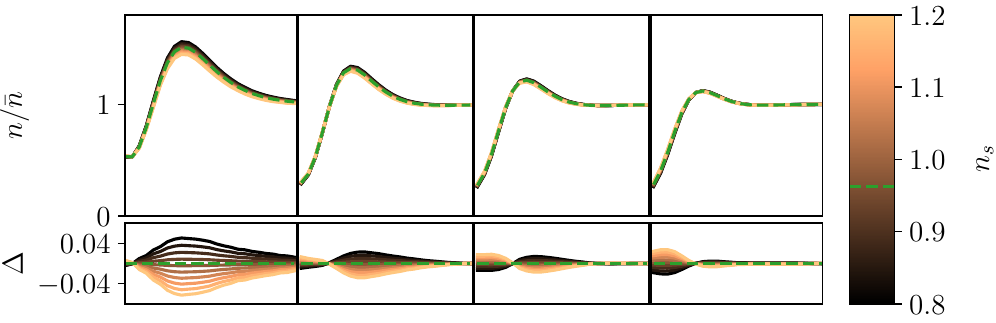}
     \end{subfigure}
     \begin{subfigure}[b]{0.49\textwidth}
    \centering
    \includegraphics[width=0.84\linewidth]{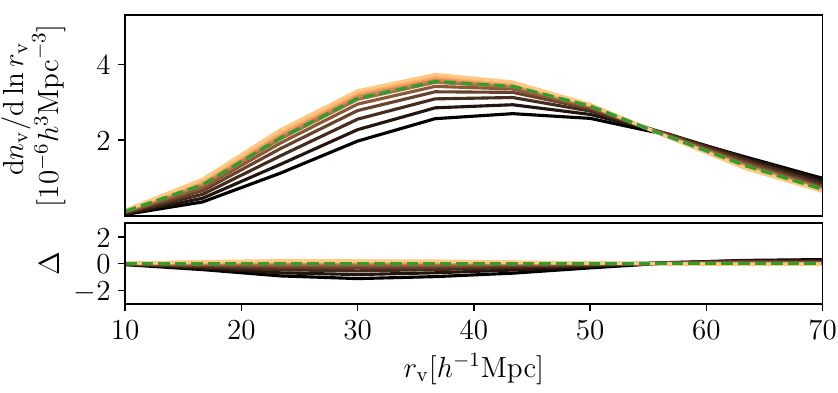}
     \end{subfigure}
     \begin{subfigure}[b]{0.49\textwidth}
    \centering
    \includegraphics[width=\linewidth]{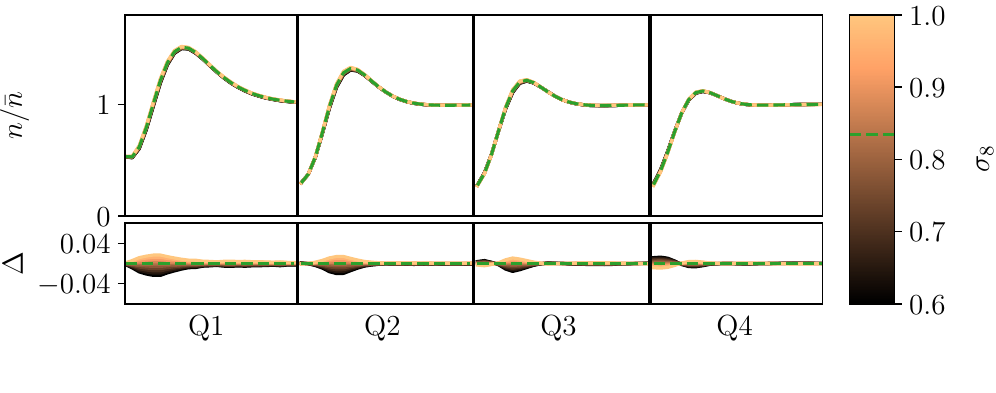}
     \end{subfigure}
    \caption{Emulator predictions of the VSF (left) and density profiles (right), varying one input parameter at a time (rightmost panels, from top to bottom). Other parameters are kept at their fiducial values for \textsc{Quijote} (see Table~\ref{tab:sim_params}). The green dashed line shows the prediction for all input parameters at their fiducial values. We show the profiles of all void size bins in each panel, where Q1 denotes the bin containing the smallest voids and Q4 the bin containing the largest voids. Lower panels show the difference of the predictions with respect to the prediction at fiducial parameter values.}
    \label{fig:emu_preds}
\end{figure*}

To obtain a qualitative understanding of the learned statistics, we show in Figure~\ref{fig:emu_preds} the VSF (left) and density profile predictions (right) of the emulator when only varying one parameter at a time. We show both the predictions themselves (upper panels) and their difference with respect to the prediction at fiducial input parameter values (lower panels). The green dashed line indicates this fiducial value prediction.

For the predictions of the VSF, we find that an increase in $\Omega_m$ leads to an increase in the abundance of ``smaller'' voids below $~50\ \mathrm{Mpc}/h$ and decreases the abundance of larger voids. For very small values of $\Omega_m$ we find an overall suppression of void numbers, which leads to a shift toward significantly larger voids. This follows expectations, as a smaller number of halos is identified at lower $\Omega_m$. The number of identified voids follows the number of halos, and as voids identified with VIDE occupy the complete simulation boxes, this leads to the learned shift in void sizes.

Increasing $\Omega_b$ leads to slight decreases in the formation of smaller voids, while marginally enhancing the number density of larger voids. Changes in $h$ and $n_s$ show similar effects on the VSF as $\Omega_m$, increasing the number of small voids and decreasing the number of large voids. The learned response to $\sigma_8$ is somewhat asymmetric: while small values of this parameter clearly decrease the number of smaller voids and increase the number of larger voids, an increase of $\sigma_8$ has little effect on the VSF.

Turning to the density profiles, we find distinctly different effects of the cosmological parameters on voids of separate size bins. When increasing $\Omega_m$, the profiles of the smallest bin are overall suppressed, i.e. void centers are slightly more underdense and compensation walls are smaller. In the other bins we also find less underdense centers of the voids but smaller compensation walls, so an overall flattening of the profile. A decrease in $\Omega_m$ leads to opposite effects, with steepened profiles for larger voids, as well as higher walls and higher central densities for the smallest set of voids. These effects can be understood through the effect of $\Omega_m$ on halo formation. For small $\Omega_m$, only few halos are formed, preferably in regions of high matter density. Due to the small chance of halos forming inside voids, the centers of large voids are more underdense and compensation walls are higher relative to the mean halo density $\Bar{n}$. Similar for the smallest subset of voids, which also experience higher compensation walls. We attribute the effect of their slightly higher inner densities for small $\Omega_m$ to small and overcompensated voids generally being identified in regions of higher local density (see~\citep{schuster_2023}).

For $\Omega_b$ we find an overall enhancement of the profiles of the two smallest bins, and deeper centers, as well as slightly larger compensation walls for the larger bins for increasing $\Omega_b$. The emulator has once more learned similar responses to changes in $h$ and $n_s$, where both parameters suppress the profiles for the smallest bin and flatten the entire profile for the other bins, similar to $\Omega_m$, but to a smaller degree. Increasing $\sigma_8$ enhances the profiles of the smaller voids, while it steepens the profiles of larger voids, i.e. deepens their underdense centers and enhances their compensation walls.

\subsection{The Effect of Cosmic Variance on the Emulators}

\begin{figure}
    \centering
    \includegraphics[width=\columnwidth]{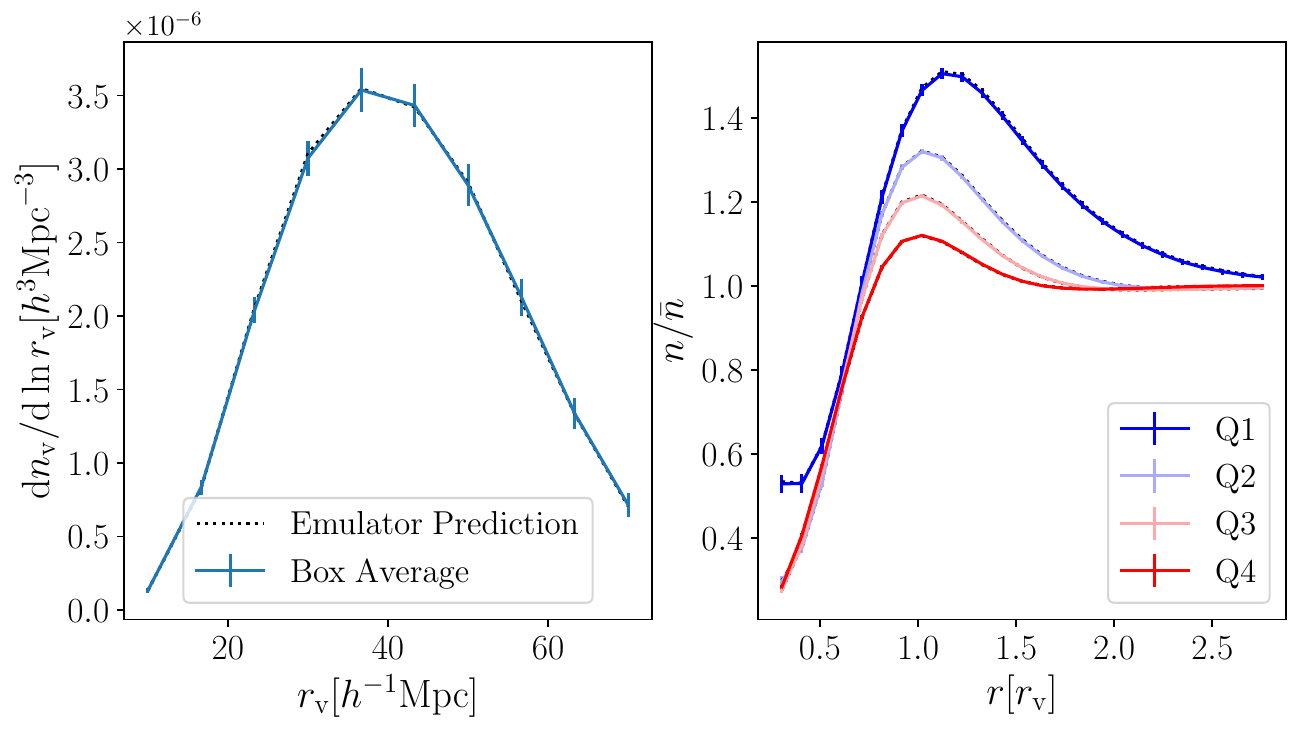}
    \caption{Predictions of the VSF (left) and the density profile emulator (right) for the fiducial cosmological parameters (dotted). For comparison, we show an average over 1000 simulation boxes (solid). `Q1' refers to the stack containing the smallest 25\% of voids and `Q4' refers to the stack containing the largest 25\% of voids.}
    \label{fig:Chap5_fiducial_test}
\end{figure}

As a consistency check, we compare the previously presented emulators' predictions to a density profile and VSF averaged over 1000 simulations at the fiducial cosmology. This means that the resulting averaged statistic should be largely unaffected by cosmic variance. Figure \ref{fig:Chap5_fiducial_test} shows this comparison along with two example boxes at the same cosmology.

The resulting prediction is very close to the averaged statistic and seems to be affected by cosmic variance to a much smaller degree. This is an expected result, as the initial conditions of the simulations are an unknown variable to the emulator. Since the initial conditions are different for every data sample, the emulator effectively marginalizes over this contribution to the data vector. 
This can also be shown explicitly by considering the optimal Bayes estimator for this problem. The neural network learns to minimize the following risk:

\begin{equation}
    \mathrm{argmin}_{\pmb{t}_\mathrm{pred}} \int (\pmb t_{\mathrm{sim}} - \pmb t_{\mathrm{pred}})\pmb C^{-1}(\pmb t_{\mathrm{sim}} - \pmb t_{\mathrm{pred}}) p(\pmb t_\mathrm{sim}|\pmb \theta) d\pmb t_\mathrm{sim},
\end{equation}

where $p(\pmb t_\mathrm{sim}|\pmb \theta)$ is the posterior of the summary statistic $\pmb t_\mathrm{sim}$ given the simulation input parameters $\pmb\theta$. Under perfect convergence the risk has a vanishing first derivative

\begin{equation}
    \frac{\partial \int (\pmb t_{\mathrm{sim}} - \pmb t_{\mathrm{pred}})\pmb C^{-1} (\pmb t_{\mathrm{sim}} - \pmb t_{\mathrm{pred}})p(\pmb t_\mathrm{sim}|\pmb \theta) d\pmb t_\mathrm{sim}}{\partial \pmb t_{\mathrm{pred}}} = 0.
\end{equation}

Via explicit calculation of the derivatives we obtain the optimal estimator

\begin{equation}
\begin{aligned}
    \pmb t_{\mathrm{pred}} &= \int p(\pmb t_\mathrm{sim}|\pmb \theta)\ \pmb t_\mathrm{sim}\ d\pmb t_\mathrm{sim}\\
    &= \int \int p(\pmb t_\mathrm{sim}|\pmb\theta,\pmb{IC})\ p(\pmb {IC})\ \pmb t_\mathrm{sim}\ d\pmb t_\mathrm{sim}\ d\pmb{IC},
\end{aligned}
\end{equation}

i.e. the posterior mean which is marginalized over the unknown initial conditions $\pmb{IC}$.

\subsection{MCMC Analysis}
Once the neural network emulators are trained, we can use them as surrogate models to predict void statistics for new input cosmological parameters. Since neural networks are extremely fast, we are able to perform an MCMC analysis to produce cosmological parameter constraints using mock data. This allows us to quantify the cosmological information content of void statistics.

We run an MCMC for mock inference, where the ``observed data'' is represented by one realization at the fiducial cosmology. The package \textsc{emcee} \citep{foreman_2013} is used for sampling, and the model prediction is provided by the emulator. The prior is set by the volume of the latin-hypercube, i.e. flat within the ranges quoted in Table~\ref{tab:sim_params}. We assume a Gaussian likelihood when comparing the mock data to the emulator-based theoretical predictions. In all contour plots presented for the emulator scheme, 16 walkers were run with $100\,000$ steps each. The first 20\% of the chains were cut to minimize burn-in effects. The covariance is estimated from 1000 fiducial realizations, and the Kaufman-Hartlap~\citep{hartlap_2007} and Dodelson-Schneider~\citep{dodelson_2013} factors are included; the resulting covariance estimate is visualized in Appendix~\ref{app:covariance}.

\subsubsection{Examining the Assumption of a Gaussian Likelihood}

\begin{figure*}
     \centering
     \begin{subfigure}[b]{0.33\textwidth}
         \centering
         \includegraphics[width=\textwidth]{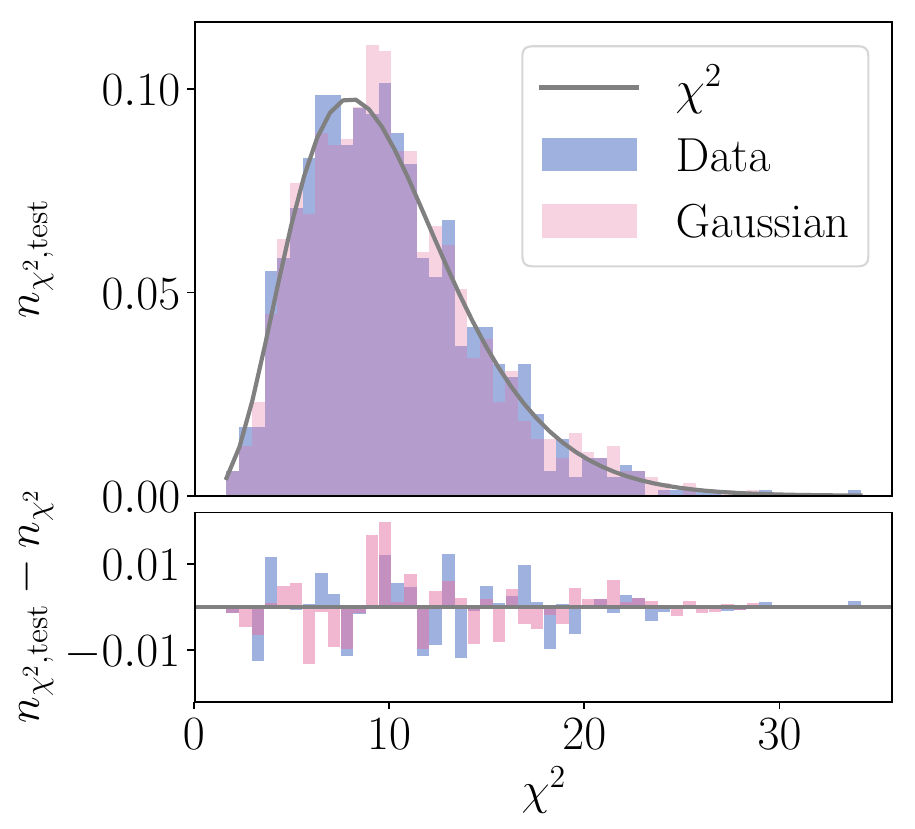}
         \caption{VSF}
     \end{subfigure}
     \begin{subfigure}[b]{0.33\textwidth}
         \centering
         \includegraphics[width=\textwidth]{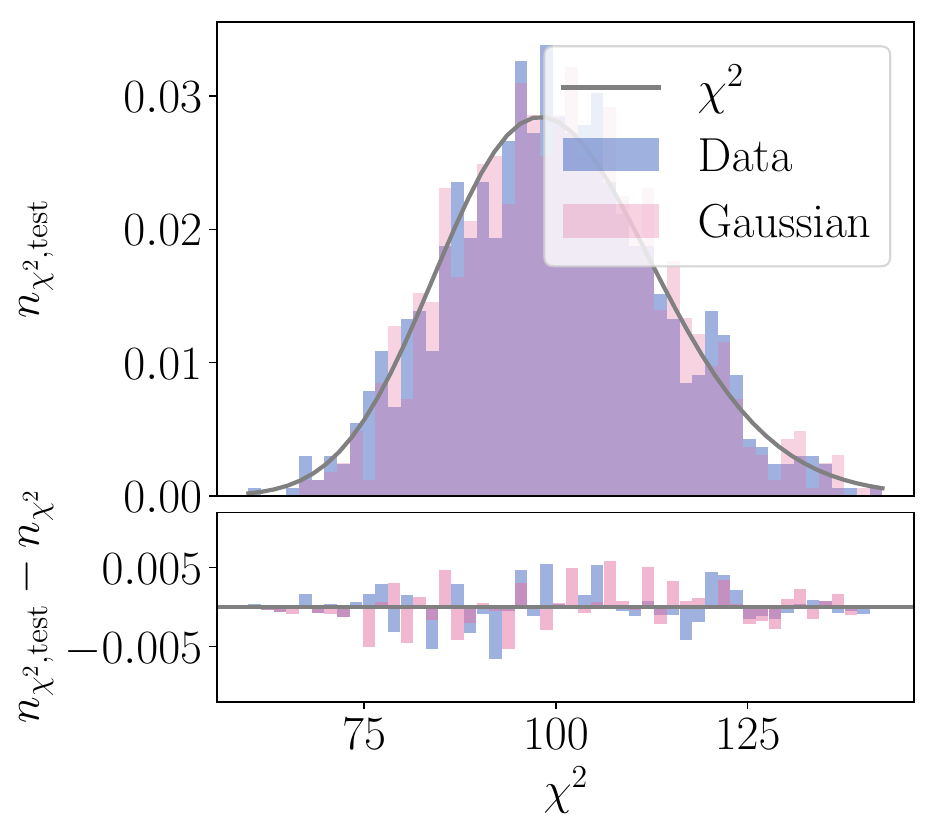}
         \caption{Density profiles}
     \end{subfigure}
     \begin{subfigure}[b]{0.33\textwidth}
         \centering
         \includegraphics[width=\textwidth]{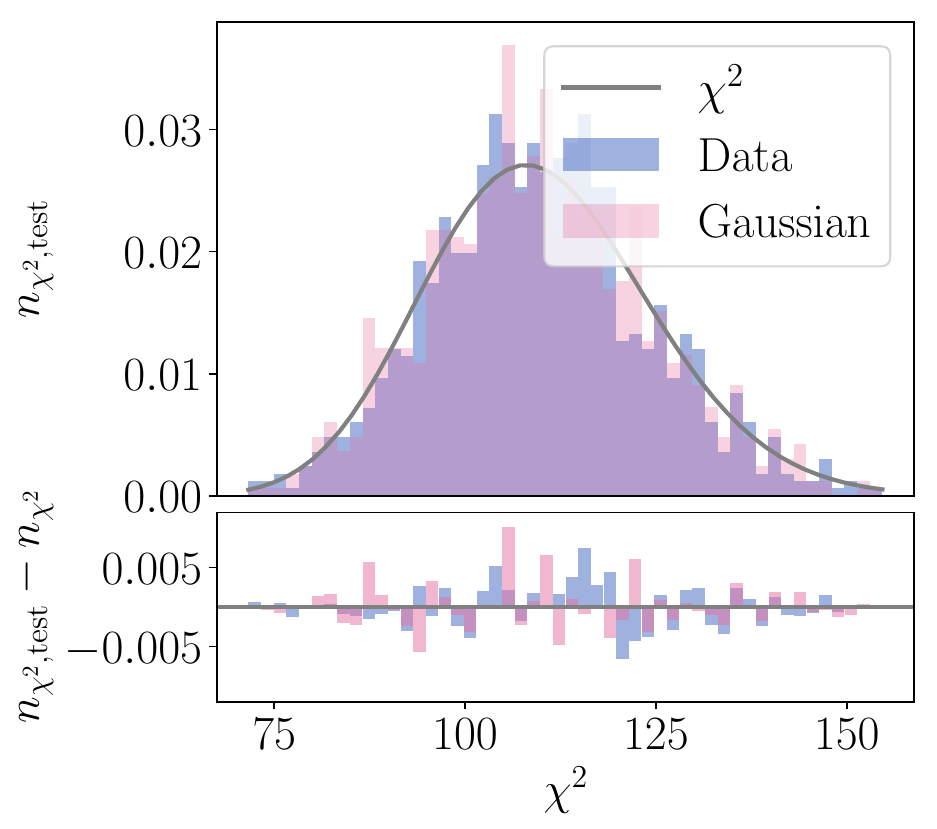}
         \caption{VSF and density profiles combined}
     \end{subfigure}
        \caption{Test of the validity of the Gaussian likelihood assumption. Each panel shows the distribution of $\chi^2$s for a different data vector in the CV set, and for a synthetic data vector drawn from a Gaussian distribution with means and standard deviations estimated from the CV set. A Gaussian data vector will result in the $\chi^2$s to be distributed according to a $\chi^2$ distribution. All three distributions are fairly close to the analytic $\chi^2$ distribution, which is an argument in favor of the Gaussianity assumption.}
        \label{fig:likelihood}
\end{figure*}

We assess the assumption of a Gaussian likelihood for the data vectors considered in this work, following \citet{friedrich_2021}. For this test we compute the $\chi^2$ between 1000 single boxes at the fiducial cosmology and their average for the VSF, the density profiles and the combination of the two. For a Gaussian likelihood, the resulting distribution of the individual $\chi^2$s will follow a $\chi^2$ distribution. To make this comparison we also draw synthetic data vectors from a Gaussian distribution around the mean data vector and the estimated covariance. We use the public code by \citet{paillas_2023}. Figure~\ref{fig:likelihood} shows the distributions of individual $\chi^2$ values, the distribution from a synthetic Gaussian data vector and a $\chi^2$ distribution with the corresponding number of degrees of freedom. The resulting distributions closely resemble the $\chi^2$ distribution, which indicates the Gaussian likelihood to be an overall reasonable approximation.

\section{Results}
\label{sec:results}
\subsection{Constraining Parameters of QUIJOTE}
\begin{figure*}
    \centering
    \includegraphics[width=.8\textwidth]{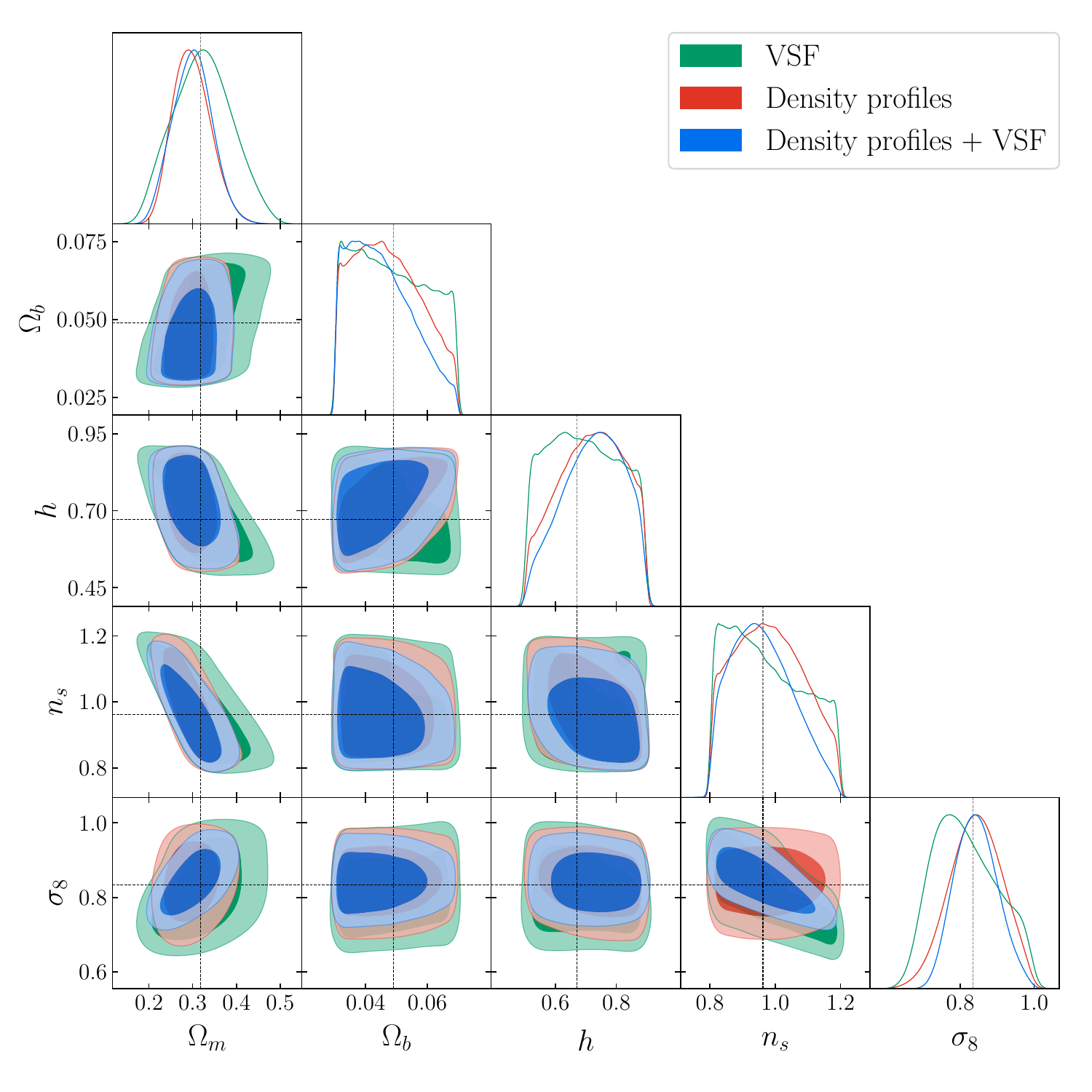}
    \caption{Forecast posterior distributions on the cosmological parameter of the simulations with the VSF, the void density profiles and their combination. $\Omega_m$ and $\sigma_8$ are constrained from void density profiles alone when running MCMC on a \textsc{Quijote} simulation at the fiducial cosmology. Including the void size function tightens contours. Contours show $1\sigma$ and $2\sigma$ credibility regions. Dashed lines indicate the true fiducial cosmology.}
    \label{fig:contours_quijote}
\end{figure*}

We now present our cosmological parameter constraints via an MCMC analysis on a single box at the fiducial cosmology, where the emulator prediction is used as the theory prediction. We note that the overall size of the box of $1h^{-1}\mathrm{Gpc}$ is still relatively small compared to actual observations. The priors on the cosmological parameters are flat within the bounds given in Table~\ref{tab:sim_params}. Figure~\ref{fig:contours_quijote} presents the forecast posteriors obtained from our emulator. We show the posterior distributions when using the VSF, the density profiles, and the combination of the two. The predictions of the density profile emulator and the VSF emulator are concatenated in order to combine the predicting power of the two observables. The resulting data vector is therefore of length $110$. In order to evaluate the likelihood, the cross covariance is estimated in the same way as the covariance was estimated previously. 

Of the five cosmological parameters, two are reasonably well constrained: $\Omega_m$ and $\sigma_8$. Other parameters remain prior dominated. Our contours are mostly in agreement with~\citet{wang_2022} who are able to additionally constrain $n_s$ from a combination of void ellipticities, density profiles and radii.~\citet{kreisch_2022} are also able to obtain weak constraints on $n_s$. However, both these works use the high resolution LH set, while we consider the fiducial resolution simulations. In comparison to the Fisher forecasts by~\citet{kreisch_2022}, we find the same degeneracies between parameters, but obtain slightly larger contours. This is consistent as the Cram\'er-Rao bound constitutes an inequality which is only reached in idealized scenarios. 

All contours are tighter when the two statistics -- the VSF and the density profiles -- are combined. Considering the degeneracies for the constrained parameters, we recover a positive $\Omega_m$-$\sigma_8$ correlation. This stems from the fact that large $\Omega_m$ values disfavor the formation of large voids and therefore have the opposite effect as large values of $\sigma_8$. For a more detailed discussion we refer to the Appendix of~\citep{contarini_2023}. The constraining power on these parameters is very promising, even though it is based on a relatively small volume of only $1h^{-1}\mathrm{Gpc}$. This is another indication of the importance of voids in unlocking complementary cosmological information in addition to the standard probes.
As the \textsc{Quijote} suite consists of gravity-only $N$-body simulations, $\Omega_b$ only influences the fluctuations in the primordial power spectrum caused by baryon acoustic oscillations. As the void profiles are equivalent to the void-halo cross correlation function~\citep{hamaus_2015}, baryon acoustic oscillations are washed out when stacking voids of different size~\citep{chan_2021}, which is why $\Omega_b$ can't be recovered. The weak constraints on $h$ from the VSF are in agreement with~\citep{contarini_2024}.

\subsection{Testing on Other Simulations} \label{sec:other_sims}

\subsubsection{Robustness to Resolution Effects}

In this section, the VSF and density profile emulators trained on the \textsc{Quijote} LH set are used to perform cosmological inference on simulations that lie out of distribution with respect to the training set regarding resolution and baryonic effects. We do so by applying our emulator based inference pipeline to a high resolution \textsc{Quijote} box and box 0 of the \texttt{Magneticum} simulations. The latter was run with higher resolution and larger box size than the training simulations, as well as a (slightly) different set of cosmological parameters. The high resolution \textsc{Quijote} simulation evolves $1024^3$ particles over the same box volume of $1h^{-1}\mathrm{Gpc}$, which is an improvement by a factor of 2 in each spatial dimension. As void statistics are sensitive to the underlying tracer density, in our case that of halos, we must reduce the halo density of the catalog obtained from this high resolution simulation. To match the halo density of the simulations used for training, we cut the halos of lowest mass from the halo catalog until we achieve the same tracer density. We choose this method instead of randomly sub-sampling halos in order to avoid the lower mass halos which are only present in the higher resolution simulation. We then perform the identical steps as before in void finding and MCMC sampling with both the VSF and density profile emulator. 

\begin{figure*}
     \centering
     \begin{subfigure}[b]{0.49\textwidth}
         \centering
        \includegraphics[width=\linewidth]{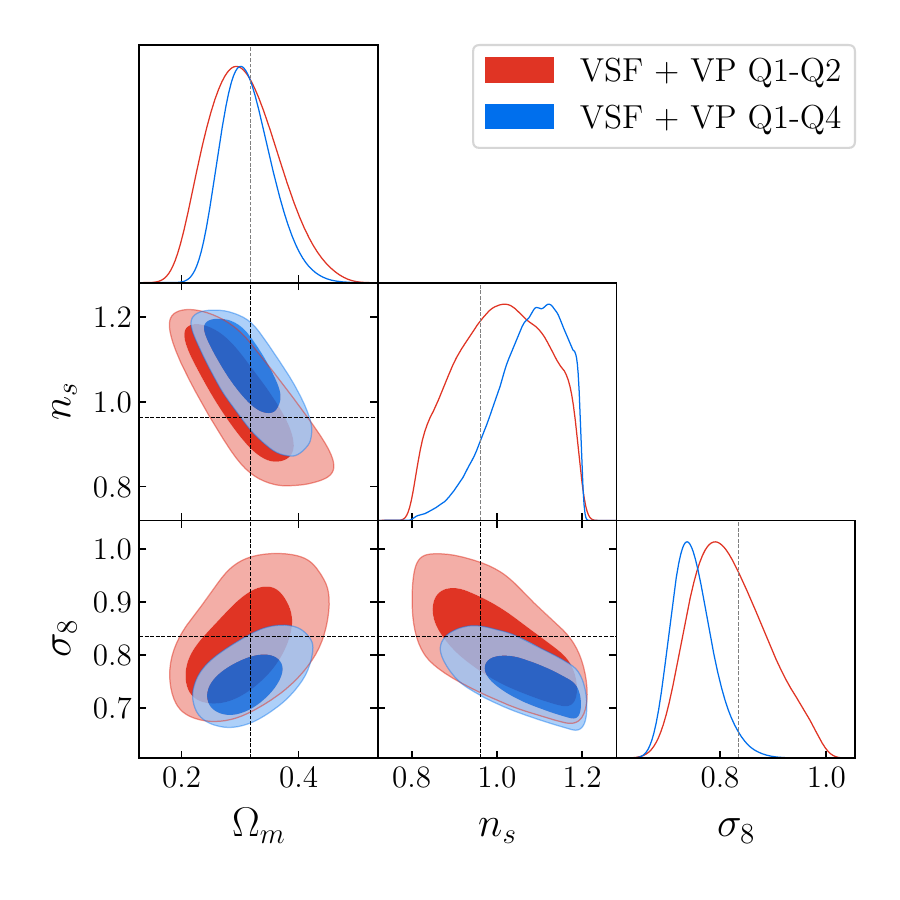}
        \caption{\textsc{Quijote} high resolution box}
        \label{fig:quijote_hr}
     \end{subfigure}
     \hfill
     \begin{subfigure}[b]{0.49\textwidth}
         \centering
         \includegraphics[width=\textwidth]{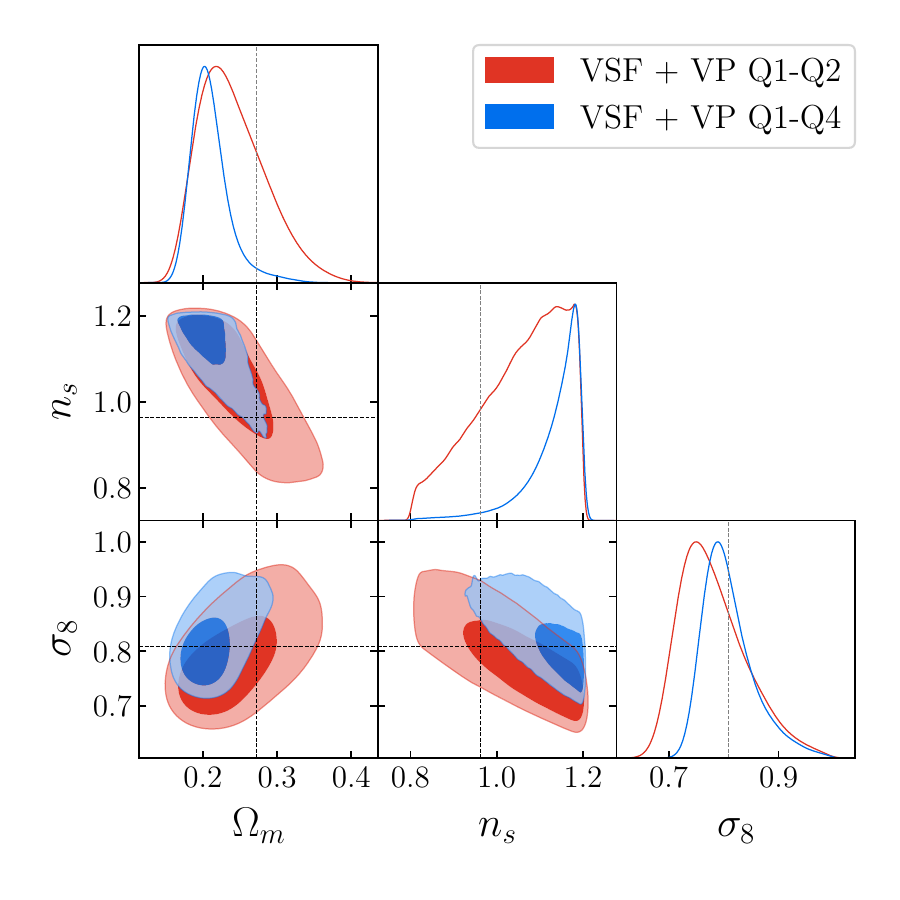}
         \caption{\texttt{Magneticum} gravity-only box 0}
         \label{fig:dmo_by_bins}
     \end{subfigure}
    \caption{Robustness tests with respect to the resolution of the simulation. Cosmological parameters of both the (a) \textsc{Quijote} high resolution box and (b) \texttt{Magneticum}  gravity-only box 0 can be inferred without bias when only smaller voids are considered. $\Omega_b$ and $h$ remain prior dominated and are not shown and marginalized over.}
\end{figure*}

We present two contours in Figure~\ref{fig:quijote_hr}, considering all void bins in the density emulator (blue) and only the smaller two bins (red), i.e. the smaller 50\% of all voids. In the case where all density profiles are used in the chain, we observe a slight bias in the $\sigma_8$ marginalized posterior. Removing the larger voids also removes this bias, albeit while also reducing the overall constraining power. We therefore find that running the MCMC with only the smaller voids yields more robust posteriors with respect to the resolution of the underlying $N$-body simulation compared to when large voids are included. We provide the following possible explanation for these findings: the mock data contains a larger number of small halos which could not be resolved in the lower resolution simulations used for training the emulator. Large voids cover the most underdense regions of the density field, which are mainly occupied by those small halos. A change in their abundance due to a shift in resolution would therefore mostly affect the density profiles of those large voids, which could explain the bias we see in the contours. We leave a further investigation up to future study.

We now move from the \textsc{Quijote} suite to the independent simulation \texttt{Magneticum}. In doing so, other than varying the resolution, we also move to a simulation box of different size. To allow for our covariance estimate to remain valid in this larger box, we cut out a $1 h^{-1}\mathrm{Gpc}$ sub-box. To handle the different resolution, in analogy to the previous section, we again reduce the tracer density of the \texttt{Magneticum} box to that of the \textsc{Quijote} simulations by omitting low-mass halos from the catalog. For the \texttt{Magneticum} simulation this is complicated by the fact that the fiducial cosmology of this simulation is slightly different; out of the entire Quijote suite, there exists no simulation with the exact same cosmological parameter values as those of \texttt{Magneticum}. Since the halo density is a function of the cosmological parameters, it is not immediately obvious which \textsc{Quijote} simulation to use as the reference density we aim to achieve. We therefore train an additional neural network, which we denote the \textit{halo number emulator}, to predict the density of halos in a \textsc{Quijote} simulation as a function of the cosmological parameters. Details on this additional neural network are given in Appendix~\ref{app:n_halo}.

We then provide the halo number emulator with the cosmological parameters of the \texttt{Magneticum} simulation. The returned value is to be interpreted as the number of halos, which a \textsc{Quijote} simulation with \texttt{Magneticum}'s cosmology would have. This value is converted into a density, which is used as the reference density. Subsequently, lower mass halos are cut from the \texttt{Magneticum} halo catalog until the reference halo density is reached. The lower mass cut determined in this way is $9.2\times10^{12}h^{-1}M_\odot$, which is comparable to the smallest halos in the \textsc{Quijote} fiducial simulations. VIDE is run on the resulting halo distribution and provides us with the final mock catalog of voids; these can then be used to measure the density profiles and the void size function to use as our final mock data. 

We now present the contours on the cosmological parameters obtained with the emulation of the void size function and the density profiles in Figure~\ref{fig:dmo_by_bins}. We again show the results when including the density profiles of all voids (blue) and when including only those of the smaller 50\% of voids (red). When including all voids, the marginal posteriors on $n_s$ and $\Omega_m$ show a small bias with respect to their true value. As with the HR \textsc{Quijote} simulation, we obtain unbiased contours once we remove the larger 50\% of voids from our analysis. This result shows that our findings regarding the impact of resolution are consistent between these two simulations. Once we correct for this by discarding the density profiles of the larger 50\% of voids, we are able to use the \textsc{Quijote} trained emulator to accurately infer $\Omega_m$ and $\sigma_8$ on another $N$-body simulation. This is very encouraging and shows the overall robustness of cosmic voids as a probe of cosmology.

\subsubsection{Robustness to Baryonic Effects}

\begin{figure}
    \centering
    \includegraphics[width=\linewidth]{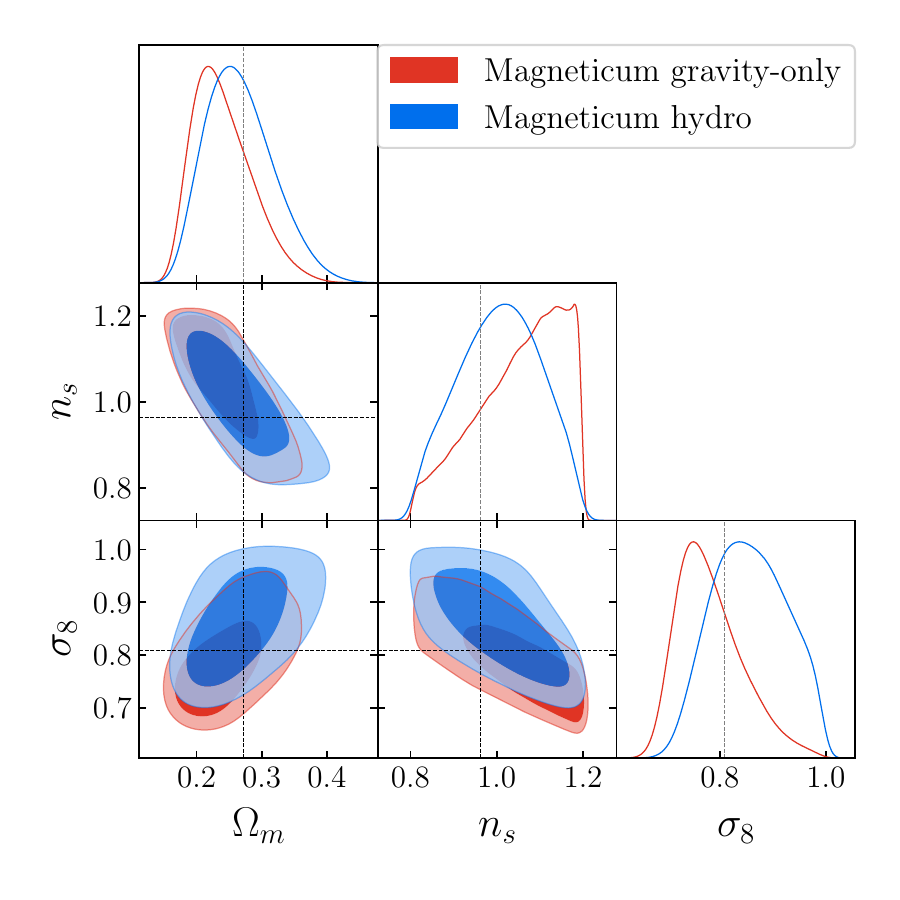}
    \caption{Posterior constraints on the \texttt{Magneticum} gravity-only (red) and hydrodynamical (blue) simulations using, in both cases, the emulator trained on \textsc{Quijote} gravity-only simulations. Using the emulators for the VSF and density profiles of the smallest 50\% of voids the parameters of the full hydrodynamic simulation \texttt{Magneticum} box 0 can be inferred. We conclude, that the voids statistics considered here are robust to baryonic effects from this simulation.}
    \label{fig:hydro}
\end{figure}

Having quantified the robustness of our approach to changes in resolution, we now examine the impact of baryonic physics on void statistics and the resulting cosmological parameters constraints. In order to be unbiased by the change in resolution, we follow the strategy presented in the previous section and only use the density profiles of the smallest 50\% of voids in our analysis while using the entire VSF. We use the full state-of-the-art hydrodynamic simulation \texttt{Magneticum} box 0, which includes implementations of both stellar and AGN feedback. Again, we infer cosmological parameters of this box, using the emulator trained solely on the \textsc{Quijote} gravity-only simulations. Results are shown in Figure~\ref{fig:hydro}. We are able to infer unbiased cosmological parameters from this hydrodynamic simulation, even though the emulator was trained on gravity-only simulations. 

We conclude that potential changes in the halo density field due to the modelling of baryonic physics in the \texttt{Magneticum} simulations are rather negligible from the perspective of common void statistics used for cosmological inference. This reaffirms previous analyses on the VSF and density profiles~\citep{schuster_2024}, where it was shown that these void statistics do not change significantly between gravity-only and hydrodynamical simulations at matched halo densities, even at considerably higher resolution than \textsc{Quijote}. Combined with our results, this is particularly interesting, as we see that voids may be a pristine probe of cosmology, irrespective of the modeling of baryonic physics in hydrodynamical simulations. This result is particularly relevant as it demonstrates that void statistics bypass the current issue of large uncertainties in baryonic feedback mechanisms. While we have only shown this explicitly here for the \texttt{Magneticum} feedback model, we would like to stress again that voids enable us to train our emulators on gravity-only simulations and subsequently perform accurate inference on a hydrodynamic simulation. We defer the study of this with regards to other hydrodynamic simulations to future work.

\section{Conclusions} \label{sec:conclusions}

We presented a set of neural network emulators to predict stacked density profiles of cosmic voids and the void size function. After performing tests on their accuracy, we performed cosmological mock inference on a \textsc{Quijote} box at the fiducial cosmology and explored the validity of our emulator approach on out-of-distribution simulations. Here we state the main results of our paper:

\begin{itemize}
    \item We trained and tested emulators for the void size function and void density profiles using a $\chi^2$ loss function. The performance of our emulators is very good with $\chi^2_\mathrm{red}$ scores of 1.08 and 2.66 respectively. The median of the individual $\chi^2_\mathrm{red}$ scores of the density profile emulator is significantly lower at 1.15, which we attribute to a few outliers. We also find 68\% of the emulator residuals to be within 1$\sigma$ as expected from cosmic variance. The presence of some outliers in the quality of predictions for density profiles at low $\Omega_m$ is due to an increase in cosmic variance in that regime.
    \item We are able to recover the parameters $\Omega_m$ and $\sigma_8$ in a \textsc{Quijote} simulation. We confirm the unique response of voids in the $\Omega_m$-$\sigma_8$ plane from their abundance, revealing their complementarity to more traditional probes of cosmology. The combination of density profiles and the VSF yields the tightest and most accurate constraints. Our analysis shows weaker constraints on the parameter $n_s$ compared to previous work due to the lower resolution of the simulations used here. It is important to note that we have not exploited any information in redshift space, as is done for the Alcock-Paczyiński test, which would add to the constraining power of voids.
    \item The robustness of our void statistic emulators is validated in terms of resolution, box size and cosmology. We use a high resolution \textsc{Quijote} simulation and a \texttt{Magneticum} simulation to construct mock data with higher resolution than that used to train the emulators. We obtain unbiased contours for the \textsc{Quijote} HR box from the smaller half of voids in that simulation. Adding large voids to our analysis biases the resulting inference, hinting at small voids being more robust to resolution effects compared to large voids. We provide a possible explanation of this with the fact that they are formed in high density environments with fewer spurious halos. We also use the \textsc{Quijote} trained emulators to recover the cosmological parameters of \texttt{Magneticum} box 0 for the gravity-only $N$-body run. In order to set a reference tracer density, a halo number density emulator is employed to predict the amount of halos in a \textsc{Quijote} simulation with the \texttt{Magneticum} cosmology. When large voids are discarded from our analysis, we can accurately recover the parameters. This shows the robustness of the emulated void statistics in regard to the underlying $N$-body simulation. 
    \item Even more notably, we also obtain unbiased contours on a void catalogue from the hydrodynamical \texttt{Magneticum} simulation when discarding larger voids. This leads us to the conclusion that small voids are robust to the \texttt{Magneticum} baryonic feedback model. This hints at an overall robustness of voids with respect to baryonic effects that warrants further study with regards to other feedback models. In light of current modeling uncertainties, the prospect of a robust probe is certainly promising if this behaviour is present in other simulations.
    \item We further presented a test of the Gaussian likelihood assumption. This is achieved by comparing the distribution of individual $\chi^2$s to an analytic $\chi^2$ distribution. The results of this test indicate that the Gaussian Likelihood assumption is valid.

\end{itemize}

This work presents an exploratory step into the response of cosmic voids to cosmological parameters and as such we leave a large amount of possible extensions to future investigation. Having examined the constraining power of VIDE halo voids in a 1$h^{-1}\mathrm{Gpc}$ volume, we propose a comparative study with other void definitions such as popcorn voids~\citep{paz_2023}, anti-halo voids~\citep{Stopyra_2020} or spherical voids. It is not clear whether different descriptions of the density field yield comparable performance in inference tasks, and therefore should be tested. Concerning the demonstrated robustness of voids to the \texttt{Magneticum} feedback model, it stands to be shown whether this is a general result or characteristic to the \texttt{Magneticum} simulations. Due to the large disagreement between different feedback implementations in state of the art simulations, this is not a direct consequence of our results.\\

\begin{acknowledgements}
We thank Carolina Cuesta-Lazaro, Oliver Friedrich, Anik Halder, Sven Krippendorf and Jochen Weller for insightful discussions. We acknowledge support via the KISS consortium (05D23WM1) funded by the German Federal Ministry of Education and Research BMBF in the ErUM-Data action plan. We are also grateful for support from the Cambridge-LMU strategic partnership. NS acknowledges support from the french government under the France 2030 investment plan, as part of the Initiative d’Excellence d’Aix-Marseille Universit\'e - A*MIDEX AMX-22-CEI-03. NS and NH acknowledge support from the Deutsche Forschungsgemeinschaft (DFG, German Research Foundation) -- HA 8752/2-1 -- 669764. The authors acknowledge additional support from the Excellence Cluster ORIGINS, which is funded by the DFG under Germany's Excellence Strategy -- EXC-2094 -- 390783311. KD acknowledges funding for the COMPLEX project from the European Research Council (ERC) under the European Union’s Horizon 2020 research and innovation program grant agreement ERC-2019-AdG 882679. The calculations for the hydrodynamical simulations were carried out at the Leibniz Supercomputing Center (LRZ) under the project pr83li. We are especially grateful for the support by M. Petkova through the Computational Center for Particle and Astrophysics (C2PAP) and for the support by N. Hammer at LRZ when carrying out the box 0 simulation within the Extreme Scale-Out Phase on the new SuperMUC Haswell extension system.
\end{acknowledgements}

%
%
\bibliographystyle{aa} 
\bibliography{bibliography} 

\appendix

\section{Covariance Matrix} \label{app:covariance}

\begin{figure}
    \centering
    \includegraphics[width=\linewidth]{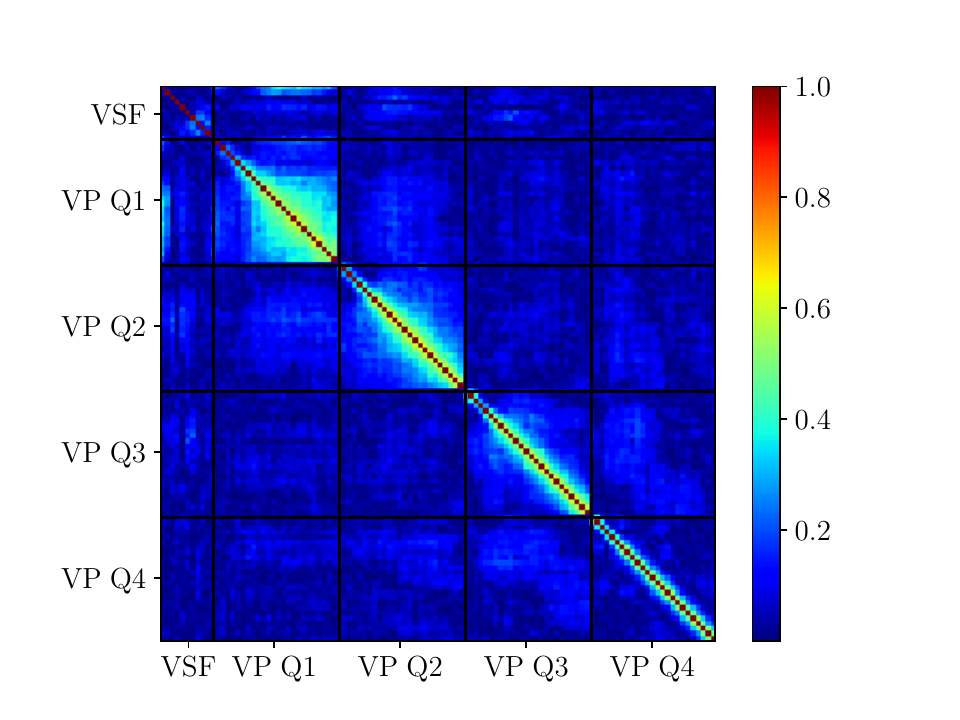}
    \caption{Absolute correlation matrix of the data vector.}
    \label{fig:correlation_matrix}
\end{figure}

We show the estimated correlation matrix of our data vector in Figure~\ref{fig:correlation_matrix}.

\section{Subsampling the Magneticum Halo Catalog} \label{app:n_halo}

\begin{figure}
    \centering
    \includegraphics[width=\linewidth]{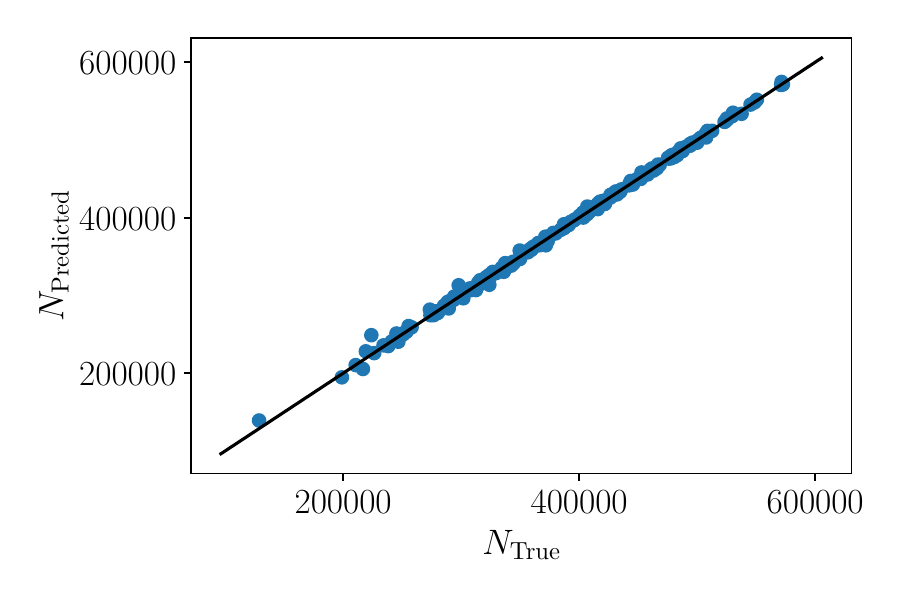}
    \caption{The performance of the halo number emulator on the test set.}
    \label{fig:n_halos}
\end{figure}

Void statistics are very sensitive to the tracer density of the underlying field, which in our case is the halo density. This in turn is a function of both cosmological parameters and the resolution of the simulation. To address this issue, we train an additional emulator to predict the number of halos in a \textsc{Quijote} box given as input the cosmological parameters. This halo number emulator consists of three hidden layers of 32 neurons each and one output neuron. Activation functions are the hyperbolic tangent for the hidden layers and a linear activation for the last layer. The input, as well as the output is standardized. The 2000 data points from the LH set are split into 1600 training, 200 validation and 200 test samples as before. The initial learning rate of the Adam solver is $10^{-4}$, and it is reduced by a factor of $10$ each time the validation loss has not improved after 10 epochs. The final learning rate is $10^{-6}$. Early stopping sets in $20$ epochs after the validation loss stopped improving and the parameters of the best epoch as measured on the validation set are restored. The mean square error, i.e.

\begin{equation}
    \mathrm{MSE} = \frac{1}{N} \sum_i (t_\mathrm{pred}-t_\mathrm{sim})^2
\end{equation}

is used as the loss function. The trained halo number emulator achieves a performance of $\mathrm{MSE} = 1.24\times 10^{-3}$ on the withheld test set. We show the predicted number density as a function of the true values for the test set in Figure~\ref{fig:n_halos}. The emulator achieves near-perfect predictions and therefore can be safely used within our pipeline without introducing additional uncertainties to our analysis.

\label{LastPage}
\end{document}